\documentclass[aps,pra,twocolumn,superscriptaddress]{revtex4-2}

\usepackage[T1]{fontenc}
\usepackage[english]{babel}

\usepackage{graphicx}
\usepackage{amsmath,amssymb}
\usepackage{multirow}

\usepackage{hyperref}

\usepackage{tikz}
\usepackage{forest}

\usepackage{braket}
\usetikzlibrary{quantikz}

\newtheorem{theorem}{Theorem}
\newtheorem{proposition}{Proposition}

\usepackage{xcolor}

\newcommand{\aDensityMatrix}{\rho}

\usepackage[backgroundcolor=orange!20, textcolor={black}, textsize=tiny]{todonotes}
\definecolor{santieditcolor}{RGB}{110,210,255}
\definecolor{cifueditcolor}{RGB}{210,210,255}
\definecolor{arieleditcolor}{RGB}{210,210,1}
\definecolor{nicoeditcolor}{RGB}{210,110,110}
\definecolor{guidoeditcolor}{RGB}{237, 145, 33}

\begin{document}

\preprint{APS/123-QED}

\title{Measuring the largest coefficients of a quantum state}
\date{\today}

\author{Nicolás Ciancaglini}
 \affiliation{Instituto de Ciencias de la Computación (ICC), CONICET-Universidad de Buenos Aires, Buenos Aires, Argentina}
\author{Santiago Cifuentes}
 \affiliation{Instituto de Ciencias de la Computación (ICC), CONICET-Universidad de Buenos Aires, Buenos Aires, Argentina}
\author{Guido Bellomo}
 \affiliation{Instituto de Ciencias de la Computación (ICC), CONICET-Universidad de Buenos Aires, Buenos Aires, Argentina}
\author{Santiago Figueira}
 \affiliation{Instituto de Ciencias de la Computación (ICC), CONICET-Universidad de Buenos Aires, Buenos Aires, Argentina}
 \affiliation{Departamento de Computación, Facultad de Ciencias Exactas y Naturales, Universidad de Buenos Aires, Buenos Aires, Argentina}
\author{Ariel Bendersky}
 \affiliation{Instituto de Ciencias de la Computación (ICC), CONICET-Universidad de Buenos Aires, Buenos Aires, Argentina}
 \affiliation{Departamento de Computación, Facultad de Ciencias Exactas y Naturales, Universidad de Buenos Aires, Buenos Aires, Argentina}

\date{\today}

\begin{abstract}
We introduce a hierarchical algorithm for identifying the indices of the largest Pauli coefficients of an unknown $n$-qubit quantum state, either those exceeding a threshold or the $t$ largest ones. The algorithm traverses a prefix-based tree whose nodes represent partial sums of squared Pauli coefficients, expanding either the nodes of largest estimated weight or all nodes above a threshold, and discarding the rest. Node weights are estimated using Bell sampling on two copies of the state, or alternatively via SWAP tests on subsystems. We prove finite-sample guarantees for the exact recovery of the target set in terms of state-dependent separation margins for the relevant prefix weights, and derive bounds on the number of nodes explored by the corresponding exact traversal. While the same Bell data can be processed exhaustively, the hierarchical strategy can reduce the classical search at the cost of potentially requiring more samples to distinguish the relevant prefix weights. The efficiency of our method depends on the number of nodes of the prefix tree with non-negligible weight as well as on the statistical conditioning of these values. We validate the method numerically on Pauli-singleton states, random stabilizer states, ground states of the transverse-field Ising model, and states with hierarchical Pauli support, illustrating different structural and statistical regimes. We further benchmark the method against alternative approaches for learning Pauli coefficients. Our work addresses the targeted discovery of dominant Pauli components without performing full state tomography.
\end{abstract}
 

\maketitle

\section{Introduction}\label{sec:intro}

The ability to characterize quantum states is crucial for many applications in quantum information processing. While several protocols exist for learning quantum states or their properties, such as \emph{quantum state tomography} \cite{Paris2004} and \emph{classical shadows} \cite{Huang2020}, a full characterization is generally inefficient due to the exponential growth of the Hilbert space dimension with the number of qubits. Consequently, most methods focus on extracting relevant information from the quantum state \cite{Aaronson2007,Ohliger2013,Chen2021,Huang2023,Wan2023,Akhtar2023} or on performing tomography for restricted families of states \cite{Gross2010,Baldwin2016,Cramer2010,Lanyon2017,Toth2010,Schreiber2025,goldar2026}.

Two-copy Bell measurements provide a particularly useful way of accessing Pauli coefficients. As shown in Ref.~\cite{Bendersky2009}, the outcomes of pairwise Bell measurements on two copies of a state can be reused to estimate squared Pauli expectation values $c_P^2=|\mathrm{Tr}(\rho P)|^2$ for arbitrary Pauli strings $P$. Thus, individual squared coefficients can be estimated without performing full state tomography. Identifying the largest coefficients by exhaustively processing all $4^n$ Pauli candidates, however, still entails exponential classical resources. This observation motivates the problem considered here: can the dominant Pauli coefficients be discovered without exhaustively processing the full set of Pauli strings? We address this question by organizing the squared coefficients into a prefix tree whose internal nodes contain the total weight of entire subsets of coefficients. Branches with sufficiently small weight can then be discarded without examining all of their descendant leaves.

Building upon these two-copy measurement ideas, this paper introduces a method for the hierarchical identification of large Pauli coefficients. Our approach identifies the indices of the $t$ largest Pauli coefficients (in descending order of magnitude when the node weights are known exactly) and, with high probability, returns exactly this set once the number of measurement samples is sufficient to resolve the relevant node decisions. Alternatively, the algorithm can identify all coefficients exceeding a specific threshold $\varepsilon$. This enables the targeted discovery of significant Pauli components without requiring full state tomography. We provide finite-sample correctness guarantees and separate two factors that control the performance of the method: the size of the relevant portion of the prefix tree, which determines the amount of classical search, and the statistical resolution required to distinguish the corresponding node weights. For instance, if the state has only $s$ nonzero nonidentity Pauli coefficients, the subtree containing these coefficients and the identity has $O(sn)$ nodes, rather than $4^n$ leaves.

Structured Pauli representations provide natural test cases for such a hierarchical search, but structure alone does not guarantee efficiency. A state may have a small or highly organized relevant subtree while the corresponding prefix weights are nevertheless difficult to distinguish statistically. Section~\ref{sec:sample_and_comp} makes this distinction precise by analyzing statistical and computational costs separately.

The underlying principle of our algorithm is the ability to efficiently estimate the sum of squared coefficients within specific subsets of $n$-qubit Pauli operators, organized as nodes of a prefix tree over the alphabet $\{I,X,Y,Z\}^n$. The algorithm always expands the node with the largest estimated weight, discarding branches unlikely to contain significant coefficients. Node weights can be estimated from Bell measurements on two copies of the state, or alternatively via SWAP measurements on subsystems.

Our approach is related to ancestral Pauli-sampling methods \cite{Hinsche2025}, which use prefix information to generate Pauli strings according to their squared-coefficient distribution, whereas our protocol uses prefix weights to search directly for dominant components. Classical shadows \cite{Huang2020} instead provide direct estimators of Pauli expectation values from single-copy measurement data, including their signs. Recent adaptive approaches for structured states \cite{goldar2026} exploit additional promises on the state family using local Pauli measurements. Specialized methods can likewise learn particular families, such as stabilizer states \cite{Montanaro2017}, more efficiently than a generic search. We compare these approaches numerically in Section~\ref{sec:simulations}.

Accordingly, the proposed tree is not intended as a uniformly sample-optimal way of processing Bell data. Its potential advantage is computational: when the dominant Pauli weight is concentrated in a small and statistically well-conditioned portion of the prefix tree, large sets of candidate operators can be discarded without explicit enumeration. Conversely, poorly resolved coarse prefix weights may require substantially more Bell samples than an exhaustive leaf-based search.

The rest of the article is organized as follows. In Section~\ref{sec:statement}, we formally define the problems of identifying large Pauli coefficients. In Section~\ref{sec:algorithm}, we describe the hierarchical algorithm and the Bell- and SWAP-based procedures used to estimate node weights. Section~\ref{sec:sample_and_comp} establishes finite-sample correctness conditions, analyzes structural and computational complexity, and compares the hierarchical strategy with exhaustive processing of Bell data and related approaches. Finally, in Section~\ref{sec:simulations}, we present numerical simulations for representative structured and many-body states and comparisons with baseline methods, followed by our conclusions in Section~\ref{sec:conclu}.

\section{Problem statement}\label{sec:statement}

One of the difficulties in identifying the dominant components of a quantum state is that its Pauli representation contains $4^n$ coefficients, while the indices of the largest ones are not known in advance. A naive exhaustive strategy must therefore address an exponentially large set of candidate observables before identifying the leading terms. We tackle this discovery problem without requiring full state tomography.

Consider an $n$-qubit density matrix $\aDensityMatrix$ with $N = 2^n$ and let
\begin{equation}
    \aDensityMatrix = \frac{1}{N} \sum_{\nu} c_{\nu} P_\nu
    \label{eq:pauli_decomposition}
\end{equation}
be its decomposition in the basis of $n$-qubit Pauli strings $P_\nu$, where $c_\nu = \mathrm{Tr}[\aDensityMatrix P_\nu]\in\mathbb{R}$. The coefficients are real because both $\aDensityMatrix$ and $P_\nu$ are Hermitian. Under this normalization, the condition $\mathrm{Tr}[\aDensityMatrix]=1$ implies $c_{I^{\otimes n}}=1$, and the coefficients satisfy
\begin{equation}
    \sum_\nu c_\nu^2 = N\,\mathrm{Tr}[\aDensityMatrix^2] \leq N,
    \label{eq:pauli_norm}
\end{equation}
with equality for pure states. Since the identity coefficient $c_{I^{\otimes n}}=1$ is fixed by normalization and is therefore known a priori, throughout the paper we exclude it from the sets of coefficients to be identified. The identity is nevertheless retained in the Pauli representation and in the node weights introduced below.

It is useful to normalize these squared coefficients as
\begin{equation}
    p_{\aDensityMatrix}(\nu) = \frac{c_\nu^2}{N\,\mathrm{Tr}[\aDensityMatrix^2]}.
    \label{eq:pauli_distribution}
\end{equation}
Eq.~\eqref{eq:pauli_norm} shows that $p_{\aDensityMatrix}$ is a probability distribution over Pauli strings.

We study the following two problems:
\begin{itemize}
    \item \textbf{P1 (threshold identification)}: Given $\varepsilon > 0$ and $\delta\in(0,1)$, identify the set
    \begin{equation}
        \mathcal{H}_\varepsilon = \{\nu\neq I^{\otimes n}: |c_\nu|>\varepsilon\}
    \label{eq:heavy_set}
    \end{equation}
    with probability at least $1-\delta$.
    \item \textbf{P2 (top-$t$ identification)}: Given $t\in\mathbb N$ and $\delta\in(0,1)$, identify with probability at least $1-\delta$ the set $\mathcal H_t$ containing the indices of the $t$ largest nonidentity Pauli coefficients in absolute value.
\end{itemize}

For later use, let $\nu_1,\nu_2,\ldots$ denote the nonidentity Pauli indices ordered so that $w_1\geq w_2\geq\cdots,w_j:=c_{\nu_j}^2$. Then
\begin{equation}
    \mathcal H_t = \{\nu_1,\ldots,\nu_t\}.
    \label{eq:top_t_set}
\end{equation}
For P2, we assume in the finite-sample analysis that the output boundary is nondegenerate, $w_t>w_{t+1}$. This condition makes $\mathcal H_t$ unique, while allowing degeneracies among coefficients that lie entirely on either side of the cutoff.

Both problems concern the identification of nonidentity Pauli indices. Since the two-copy measurements used below provide direct access to squared coefficients, the search identifies their magnitudes but not their signs. Once the relevant indices have been identified, the corresponding signed expectation values can, if needed, be estimated separately by direct Pauli measurements.

In Section~\ref{sec:algorithm} we introduce a hierarchical algorithm for both problems, show that it is correct when the node weights are known exactly, and describe how these weights are estimated from Bell and SWAP measurements. The finite-sample conditions under which the desired success probability is guaranteed are analyzed in Section~\ref{sec:sample_and_comp}.

\section{Algorithm}\label{sec:algorithm}

\subsection{Prefix-based tree structure}
A generalized Pauli operator on $n$ qubits can be identified with a string of length $n$ over the alphabet $\{I,X,Y,Z\}$. Given two strings $\mu$ and $\sigma$, we denote their concatenation by $\mu\sigma$, and we say that $\mu$ is a prefix of $\nu$ if $\nu$ begins with $\mu$. As usual, $\lambda$ denotes the empty string.

For each prefix $\mu$, we define a node in a tree
\begin{equation}
    \Gamma_\mu = \{\nu\in\{I,X,Y,Z\}^n:\mu\text{ is a prefix of }\nu\},
\end{equation}
and associate with this node the weight
\begin{equation}
    K_\mu = \sum_{\nu\in\Gamma_\mu}c_\nu^2 .
    \label{eq:node_weight}
\end{equation}
The root is $\Gamma_\lambda$, with
\begin{equation}
    K_\lambda = N\,\mathrm{Tr}[\aDensityMatrix^2],
\end{equation}
whereas at a leaf $\nu$ we have $K_\nu=c_\nu^2$. This tree structure is shown in Figure \ref{fig:tree2}.

Every internal node $\Gamma_\mu$, with $|\mu|<n$, has four children, $\Gamma_{\mu I}$, $\Gamma_{\mu X}$, $\Gamma_{\mu Y}$, and $\Gamma_{\mu Z}$. These sets form a disjoint partition of $\Gamma_\mu$, and consequently
\begin{equation}
    K_\mu = K_{\mu I} + K_{\mu X} + K_{\mu Y} + K_{\mu Z}.
    \label{eq:node_additivity}
\end{equation}

\begin{figure*}[ht]
    \begin{center}
        \fbox{\begin{forest}
            [$\Gamma_\lambda$
            [$\Gamma_I$
                [$\Gamma_{II}$]
                [$\Gamma_{IX}$]
                [$\Gamma_{IY}$]
                [$\Gamma_{IZ}$]]
            [$\Gamma_X$[][][][]]
            [$\Gamma_Y$[][][][]]
            [$\Gamma_Z$[][][][]]
            ]
        \end{forest}}
    \end{center}
    \caption{Prefix tree for $n$-qubit Pauli strings. Each internal node $\Gamma_\mu$ is partitioned into the four subsets obtained by appending $I$, $X$, $Y$, or $Z$ to the prefix.}
    \label{fig:tree2}
\end{figure*}

Using Eq.~\eqref{eq:pauli_distribution}, the normalized node weight is
\begin{equation}
    \frac{K_\mu}
    {N\,\mathrm{Tr}[\aDensityMatrix^2]} = \sum_{\nu\in\Gamma_\mu} p_{\aDensityMatrix}(\nu).
    \label{eq:prefix_marginal}
\end{equation}
Thus, $K_\mu$ is proportional to the probability mass carried by the prefix $\mu$ under the squared-Pauli distribution.

The algorithm maintains a set of nodes, the \emph{frontier}, which initially contains only the root. At each step it selects a node of the frontier and removes it. If the node is internal, its four children are added to the frontier; if it is a leaf, it is returned, unless it is the known identity operator, which is discarded. Since the children of a node partition it, at every step the sets $\Gamma_\mu$ of the frontier nodes, together with the leaves already returned or discarded, form a partition of $\{I,X,Y,Z\}^n$. We first assume that the weights $K_\mu$ are known exactly; in Section~\ref{sec:sample_and_comp} the same procedures are run with the estimators $\hat K_\mu$ introduced below. The two problems differ in the rule that selects the node and in when the procedure stops:
\begin{itemize}
    \item \textbf{P1 (threshold traversal):} select any frontier node with weight above $\varepsilon^2$; stop when there is none. The remaining nodes are never selected, and are said to be \emph{pruned}.
    \item \textbf{P2 (best-first traversal):} select a frontier node of maximum weight; stop after $t$ nonidentity leaves have been returned.
\end{itemize}

We first establish correctness of the best-first traversal for P2. With exact weights, the best-first traversal returns leaves in nonincreasing order of $c_\nu^2$. Indeed, when a leaf $\nu$ is selected, the partition property implies that every other leaf $\xi$ not yet returned lies in $\Gamma_\mu$ for some frontier node $\mu$, so
\begin{equation}
    c_\xi^2 \leq K_\mu \leq K_\nu = c_\nu^2,
\end{equation}
because $K_\nu$ is maximal on the frontier. Discarding the known identity leaf preserves this ordering among the nonidentity leaves, so the first $t$ returned leaves form $\mathcal H_t$.

The threshold traversal is also exact. If a target leaf $\nu$ satisfies $|c_\nu|>\varepsilon$, then every ancestor $\mu$ of $\nu$ satisfies
\begin{equation}
    K_\mu \geq c_\nu^2 > \varepsilon^2,
\end{equation}
so its branch is never pruned, and since the procedure stops only when no node above the threshold remains, $\nu$ is eventually returned. Conversely, a leaf with $|c_\nu|\leq\varepsilon$ has $K_\nu\leq\varepsilon^2$ and is never selected.

The two arguments above prove the following.
\begin{proposition}[exact traversal]\label{prop:exact}
Assume the node weights $K_\mu$ are known exactly. Then (i) the best-first traversal for \emph{P2} returns the nonidentity leaves in nonincreasing order of $c_\nu^2$, and the first $t$ of them form $\mathcal H_t$ (unique if $w_t>w_{t+1}$); (ii) the threshold traversal for \emph{P1} returns exactly $\mathcal H_\varepsilon$.
\end{proposition}

As an illustration, consider the two-qubit state
\begin{equation}
\rho=\tfrac14\left(I\otimes I+0.6\,X\otimes X+0.4\,Z\otimes Z\right),
\label{eq:example}
\end{equation}
whose nonzero coefficients are $c_{II}=1$, $c_{XX}=0.6$ and $c_{ZZ}=0.4$, so that $K_\lambda=N\,\mathrm{Tr}[\rho^2]=1.52$. The four children of the root have weights $K_I=1$, $K_X=0.36$, $K_Y=0$ and $K_Z=0.16$, which add up to $K_\lambda$ as in Eq.~\eqref{eq:node_additivity}. For P1 with $\varepsilon=1/2$, only $\Gamma_I$ and $\Gamma_X$ exceed $\varepsilon^2=0.25$ and are expanded: the leaf $II$ is discarded as the identity, the leaf $XX$ is returned, and the eight leaves below $\Gamma_Y$ and $\Gamma_Z$ are never examined. For P2 with $t=2$, the traversal expands $\Gamma_I$ (weight $1$) and discards $II$, expands $\Gamma_X$ (weight $0.36$) and returns $XX$, then expands $\Gamma_Z$ (weight $0.16$) and returns $ZZ$; thus $\mathcal{H}_2=\{XX,ZZ\}$ is obtained in decreasing order of $c_\nu^2$, and $\Gamma_Y$ is never expanded.

The structural and finite-sample costs of the traversal are analyzed in Section~\ref{sec:sample_and_comp}.

\subsection{Estimating node values}

The node weights admit a useful operator representation that directly connects Bell sampling and SWAP measurements. To derive it, we use the standard Pauli expansion of the two-qubit SWAP operator,
\begin{equation}
    \mathrm{SWAP}
    =
    \frac{1}{2}\left(I\otimes I + X\otimes X + Y\otimes Y + Z\otimes Z\right).
    \label{eq:pauli_swap_identity}
\end{equation}

Let $\mu$ be a prefix of length $k$, and denote by $A$ the first $k$ qubits and by $B$ the remaining $n-k$ qubits. Every $\nu\in\Gamma_\mu$ can be written as $\nu=\mu\beta$, where $\beta$ runs over all Pauli strings on $B$. Denoting the two copies by $1$ and $2$, we therefore have
\begin{align}
    \sum_{\nu\in\Gamma_\mu}
    P_\nu^{(1)}\otimes P_\nu^{(2)}
    &=
    \left(
    P_\mu^{(A_1)}\otimes P_\mu^{(A_2)}
    \right)
    \sum_{\beta}
    P_\beta^{(B_1)}\otimes P_\beta^{(B_2)}
    \nonumber\\
    &=
    2^{n-k}
    \left(
    P_\mu^{(A_1)}\otimes P_\mu^{(A_2)}
    \right)
    \mathrm{SWAP}_B .
    \label{eq:prefix_swap_identity}
\end{align}
We therefore define
\begin{equation}
    U_\mu = \left( P_\mu^{(A_1)}\otimes P_\mu^{(A_2)} \right) \mathrm{SWAP}_B,
    \label{eq:node_observable_definition}
\end{equation}
where $\mathrm{SWAP}_B$ exchanges subsystems $B_1$ and $B_2$. Since
\begin{equation}
    c_\nu^2 = \mathrm{Tr} \left[(\aDensityMatrix\otimes\aDensityMatrix)(P_\nu^{(1)}\otimes P_\nu^{(2)})\right],
\end{equation}
summing over $\Gamma_\mu$ yields
\begin{equation}
    K_\mu = 2^{n-k} \langle U_\mu\rangle_{\aDensityMatrix\otimes\aDensityMatrix}.
    \label{eq:node_observable}
\end{equation}
Thus, estimating a node weight reduces to estimating the expectation value of the Hermitian unitary $U_\mu$, whose eigenvalues are $\{-1,+1\}$.

\paragraph*{Bell sampling.}
Bell sampling consists of measuring each pair of corresponding qubits, one from each copy, in the Bell basis; we index the pairs by $r=0,\dots,n-1$. The Bell basis simultaneously diagonalizes $X\otimes X$, $Y\otimes Y$, $Z\otimes Z$ and $\mathrm{SWAP}$. The eigenvalues of $(X\otimes X,\,Y\otimes Y,\,Z\otimes Z)$ are $(-,+,+)$, $(+,-,+)$ and $(+,+,-)$ on $|\Phi^-\rangle$, $|\Phi^+\rangle$ and $|\Psi^+\rangle$ respectively, and $(-,-,-)$ on the singlet $|\Psi^-\rangle=(|01\rangle-|10\rangle)/\sqrt2$; $I\otimes I$ always has eigenvalue $+1$, and $\mathrm{SWAP}$ has eigenvalue $-1$ on the singlet and $+1$ on the other three states. Since $P_\mu\otimes P_\mu$ and $\mathrm{SWAP}_B$ factorize over pairs, $U_\mu$ is diagonal in the Bell basis of the $n$ pairs, so
the outcomes of a single Bell-measurement run $j$ determine its eigenvalue for every prefix $\mu$ simultaneously.

In run $j$, let $s_{j,r}(\sigma)\in\{-1,+1\}$ be the eigenvalue of $\sigma\otimes\sigma$ on pair $r$, and let $a_{r,j}=1$ if pair $r$ is found in the singlet and $a_{r,j}=0$ otherwise. For a prefix $\mu$ of length $k$ define
\begin{equation}
    s_j(\mu)=\prod_{r=0}^{k-1}s_{j,r}(\mu_r),
    \qquad
    m_{k,j}=\sum_{r=k}^{n-1}a_{r,j}.
\end{equation}
Here $s_j(\mu)$ is the eigenvalue of $P_\mu\otimes P_\mu$ on the first $k$ pairs (the product of the single-pair eigenvalues, because that operator is a tensor product over pairs), and $m_{k,j}$ is the number of singlets among the remaining pairs; only its parity matters, because $\mathrm{SWAP}_B$ contributes a factor $-1$ for each singlet. Therefore the eigenvalue of $U_\mu$ in run $j$ is
\begin{equation}
    u_{\mu,j}=s_j(\mu)\,(-1)^{m_{k,j}}\in\{-1,+1\}.
    \label{eq:node_run_variable}
\end{equation}

Eq.~\eqref{eq:node_observable} therefore gives the unbiased estimator
\begin{equation}
    \hat K_\mu = \frac{2^{n-k}}{M} \sum_{j=1}^{M}u_{\mu,j},
    \label{eq:bell_K_estimator}
\end{equation}
where $M$ is the number of independent Bell repetitions, each consuming two copies of $\rho$. For a fixed node the variables $u_{\mu,j}$, $j=1,\dots,M$, are independent and bounded, which is what allows Hoeffding's inequality to be applied in Section~\ref{sec:sample_and_comp}. An alternative derivation of Eq.~\eqref{eq:bell_K_estimator}, directly from the sign patterns of the Bell-measurement outcomes, is given in Appendix~\ref{app:bell_estimator}. The single-run variable $u_{\mu,j}$ has expectation
\begin{equation}
    \mathbb E[u_{\mu,j}] = \frac{K_\mu}{2^{n-k}},
    \label{eq:node_normalized_mean}
\end{equation}
and hence
\begin{equation}
    \mathrm{Var}(\hat K_\mu) = \frac{2^{2(n-k)}-K_\mu^2}{M}.
    \label{eq:node_variance}
\end{equation}
Although $K_\mu\geq0$, the finite-sample estimator $\hat K_\mu$ need not be nonnegative.

Two limiting cases are instructive. At a leaf, $k=n$, there is no suffix and $\hat K_\nu=\frac1M\sum_j s_j(\nu)$ is the estimator of $c_\nu^2$ of Ref.~\cite{Bendersky2009}; for a general prefix, the factor $(-1)^{m_{k,j}}$ accounts for the suffix. At the root, $k=0$, $\hat K_\lambda$ estimates $N\,\mathrm{Tr}[\rho^2]$.

The same Bell record can be reused to evaluate any node subsequently queried by the tree. Moreover, the run-dependent variables of the children of a node follow directly
from those of the parent. If $\sigma\in\{I,X,Y,Z\}$ is appended to a prefix $\mu$
of length $k$, then $s_j(\mu\sigma)=s_j(\mu)\,s_{j,k}(\sigma)$ and
$m_{k+1,j}=m_{k,j}-a_{k,j}$, so
\begin{equation}
    u_{\mu\sigma,j}=u_{\mu,j}\,s_{j,k}(\sigma)\,(-1)^{a_{k,j}}.
    \label{eq:child_update}
\end{equation}
Since $s_{j,k}(\sigma)$ and $a_{k,j}$ are read directly from the Bell record, each
child vector $\{u_{\mu\sigma,j}\}_{j=1}^M$ is obtained from the parent's in $O(M)$
operations.

It follows in particular that, for every run $j$,
\begin{equation}
    \sum_{\sigma\in\{I,X,Y,Z\}}u_{\mu\sigma,j}
    =u_{\mu,j}\,(-1)^{a_{k,j}}\sum_{\sigma}s_{j,k}(\sigma)
    =2u_{\mu,j},
    \label{eq:run_child_additivity}
\end{equation}
because $\sum_\sigma s_{j,k}(\sigma)=+2$ if pair $k$ is not the singlet (one of the four eigenvalues is $-1$ and three are $+1$) and $-2$ if it is (the eigenvalues of $X,Y,Z$ are $-1$ and that of $I$ is $+1$). Using the different prefactors in Eq.~\eqref{eq:bell_K_estimator} for depths $k$ and $k+1$, this gives
\begin{equation}
    \hat K_\mu=\hat K_{\mu I}+\hat K_{\mu X}+\hat K_{\mu Y}+\hat K_{\mu Z}
    \label{eq:estimator_additivity}
\end{equation}
exactly for every finite Bell-measurement record, not only in expectation. Unlike the exact weights, however, the estimates need not decrease along a branch: since $\hat K_\mu$ can be negative, a child may exceed its parent. The finite-sample analysis of Section~\ref{sec:sample_and_comp} therefore does not rely on monotonicity.

\paragraph*{SWAP implementation.}

Eq.~\eqref{eq:node_observable} also provides a direct SWAP-based implementation. The observable $U_\mu$ can be measured with a Hadamard test on two copies of $\rho$, controlling the prefix-dependent Pauli operators and the SWAP on subsystem $B$ (Figure~\ref{fig:circuit}). Let $z_{\mu,j}\in\{-1,+1\}$ denote the ancilla outcome in repetition $j$. Since $U_\mu$ is Hermitian, $\mathbb E[z_{\mu,j}]=\langle U_\mu\rangle$, and by
Eq.~\eqref{eq:node_observable}
\begin{equation}
    \mathbb E[z_{\mu,j}] = 2^{k-n}K_\mu,
\end{equation}
and therefore
\begin{equation}
    \hat K_\mu^{\mathrm{SWAP}} = \frac{2^{n-k}}{M} \sum_{j=1}^{M}z_{\mu,j}
\end{equation}
is an unbiased estimator of $K_\mu$.

\begin{figure}
    \centering
    \begin{quantikz}
    \lstick{$\lvert 0\rangle$} & \gate{H} & \ctrl{3} & \ctrl{4} & \gate{H} & \meter{} \\
\lstick[2]{$\rho$} &   \qw  & \gate{P_{\mu}}  & \qw  & \qw & \qw \\
& \qw & \qw & \swap{2} & \qw & \qw\\
\lstick[2]{$\rho$} &   \qw   & \gate{P_{\mu}}  & \qw  & \qw & \qw \\
& \qw & \qw & \targX{} & \qw & \qw
    \end{quantikz}
    \caption{Circuit used to estimate $K_{\mu}$ using a modification of the SWAP test.}
    \label{fig:circuit}
\end{figure}

Both procedures measure the same binary observable $U_\mu$, and therefore have the same fixed-node outcome distribution. Operationally, however, they differ in an important way: the SWAP-based measurement circuit depends explicitly on the prefix $\mu$, whereas a single Bell-measurement record can be reused to estimate arbitrary nodes encountered during the traversal. For this reason, the resource analysis in Section~\ref{sec:sample_and_comp} focuses primarily on the Bell implementation.

\section{Sample and computational efficiency}
\label{sec:sample_and_comp}

The finite-sample analysis separates two distinct questions. First, how
many Bell samples are required to guarantee that the correct coefficients
are returned? Second, how many nodes must be processed in order to find
them? Statistical fluctuations may cause the finite-sample algorithm to
explore additional branches without changing its final output. We therefore
analyze correctness and computational cost separately.

\paragraph*{Uniform finite-sample bound.}

From Eq.~\eqref{eq:bell_K_estimator}, the estimator of a node value is a
rescaled average of independent random variables taking values in
$\{-1,+1\}$. Therefore, for any fixed node $\Gamma_\mu$ and any $\eta>0$,
Hoeffding's inequality gives
\begin{equation}
    \Pr\!\left[
    |\hat K_\mu-K_\mu|
    >
    2^{n-|\mu|}\eta
    \right]
    \leq
    2e^{-M\eta^2/2}.
    \label{eq:node_hoeffding}
\end{equation}
The factor $2^{n-|\mu|}$ reflects the depth-dependent statistical scale
already apparent in the exact variance of Eq.~\eqref{eq:node_variance}.

We now choose $\eta$ so that this bound holds simultaneously for the
complete prefix tree, which contains
\begin{equation}
    L_n
    =
    \sum_{k=0}^{n}4^k
    =
    \frac{4^{n+1}-1}{3}
\end{equation}
nodes. Defining
\begin{equation}
    \eta_M(\delta)
    =
    \sqrt{
    \frac{2}{M}
    \log\!\left(\frac{2L_n}{\delta}\right)
    },
    \label{eq:eta_M}
\end{equation}
a union bound implies that, with probability at least $1-\delta$,
\begin{equation}
    |\hat K_\mu-K_\mu|
    \leq
    2^{n-|\mu|}\eta_M(\delta)
    \qquad
    \text{for every node }\mu .
    \label{eq:uniform_node_bound}
\end{equation}
Thus, $\eta_M(\delta)$ is the uniform statistical resolution supplied by
$M$ Bell samples, at confidence level $1-\delta$, after accounting for the
depth-dependent scale of the node estimator. Note that the node estimators need not be independent for this union bound to hold;
in the Bell implementation they are in fact obtained from the same
measurement record.

Eq.~\eqref{eq:uniform_node_bound} does not by itself make near-root estimates accurate: their absolute error is still of order $2^{n-|\mu|}\eta_M(\delta)$. Its role is to fix a single resolution $\eta_M(\delta)$ which, in Theorems~\ref{thm:P1} and \ref{thm:P2} below, is compared with state-dependent margins measured on the same depth-dependent scale.

\paragraph*{Exhaustive processing of Bell data.}

Before analyzing the hierarchical search, it is useful to consider what
can be obtained by processing the same Bell data directly at the leaves.
As discussed in Ref.~\cite{Bendersky2009}, a Bell-measurement record can
be reused to estimate squared Pauli expectation values for arbitrary Pauli
strings. At a leaf, $|\mu|=n$, the depth-dependent factor in
Eq.~\eqref{eq:uniform_node_bound} is equal to one.

For P1, define the minimum leaf separation from the threshold as
\begin{equation}
    \Delta_\varepsilon
    =
    \min_{\nu\neq I^{\otimes n}}
    |c_\nu^2-\varepsilon^2|.
    \label{eq:leaf_gap_p1}
\end{equation}
If $\Delta_\varepsilon>0$, we choose $\eta_M(\delta)=\Delta_\epsilon$. A union bound over all nonidentity leaves shows that exhaustive processing correctly classifies every coefficient with probability at least $1-\delta$ whenever
\begin{equation}
    M
    >
    \frac{2}{\Delta_\varepsilon^2}
    \log\!\left(
    \frac{2(4^n-1)}{\delta}
    \right).
    \label{eq:exhaustive_p1}
\end{equation}
Thus, for fixed $\Delta_\varepsilon$,
\begin{equation}
    M
    =
    O\!\left(
    \frac{n+\log(1/\delta)}{\Delta_\varepsilon^2}
    \right).
\end{equation}

For P2, recall the assumption $w_t>w_{t+1}$ and define the midpoint
\begin{equation}
    \theta_t
    =
    \frac{w_t+w_{t+1}}{2}.
    \label{eq:theta_t}
\end{equation}
The distance from $\theta_t$ to either side of the output boundary is
$(w_t-w_{t+1})/2$. Hence exhaustive processing separates the $t$ target
leaves from all remaining nonidentity leaves with probability at least
$1-\delta$ whenever
\begin{equation}
    M
    >
    \frac{8}{(w_t-w_{t+1})^2}
    \log\!\left(
    \frac{2(4^n-1)}{\delta}
    \right).
    \label{eq:exhaustive_p2}
\end{equation}

Thus, for fixed leaf gaps, Bell sampling itself has favorable sample
complexity. The drawback of exhaustive processing is classical: explicitly
searching the complete set of Pauli candidates requires handling an
exponentially large set. The hierarchical strategy aims to avoid this
enumeration by using coarser prefix weights, whose statistical resolution
may instead become the limiting resource.

\paragraph*{Threshold identification (P1).}

For P1 we only need two things: that no branch containing a coefficient above the threshold is pruned, and that no leaf below the threshold is returned. The sampled traversal may otherwise differ from the exact one.

We write $\mu\preceq\nu$ when $\mu$ is a prefix of $\nu$, including the empty prefix and $\nu$ itself. Define
\begin{equation}
    \gamma_\varepsilon
    =
    \min\left\{
    \Delta_\varepsilon,
    \;
    \min_{\substack{\nu\in\mathcal H_\varepsilon\\
                     \mu\preceq\nu}}
    \frac{K_\mu-\varepsilon^2}{2^{n-|\mu|}}
    \right\},
    \label{eq:gamma_epsilon}
\end{equation}
where the second term is omitted if $\mathcal H_\varepsilon=\varnothing$. For every target leaf $\nu\in\mathcal H_\varepsilon$ and every $\mu\preceq\nu$,
\begin{equation}
    K_\mu\geq c_\nu^2>\varepsilon^2,
\end{equation}
so the second term in Eq.~\eqref{eq:gamma_epsilon}, whenever present, is strictly positive. Consequently, $\gamma_\varepsilon>0$ if and only if no nonidentity Pauli coefficient lies exactly at the threshold, $|c_\nu|=\varepsilon$.

Intuitively, $\gamma_\varepsilon$ is the smallest margin, measured on the scale of each depth, by which a decision that matters must be resolved: keeping a prefix of a target leaf, or rejecting a non-target leaf.

If
\begin{equation}
    \eta_M(\delta)<\gamma_\varepsilon ,
    \label{eq:p1_condition}
\end{equation}
then Eq.~\eqref{eq:uniform_node_bound} and the definition of $\gamma_\varepsilon$ imply that every prefix $\mu$ of a target leaf satisfies
\begin{equation}
    |\hat K_\mu-K_\mu|
    \leq 2^{n-|\mu|}\eta_M(\delta)
    < 2^{n-|\mu|}\gamma_\varepsilon
    \leq K_\mu-\varepsilon^2 ,
\end{equation}
and therefore $\hat K_\mu>\varepsilon^2$.

Thus no target branch can be pruned. Moreover, since $\gamma_\varepsilon\leq\Delta_\varepsilon$, every non-target leaf remains
below the threshold and cannot be returned. Finally, as long as a target leaf remains unrecovered, one of its prefixes remains in the frontier with estimated weight above $\varepsilon^2$, so the traversal cannot terminate; since the tree is finite, all target leaves are eventually returned. The finite-sample algorithm therefore returns exactly $\mathcal H_\varepsilon$ with probability at least $1-\delta$.

Using Eq.~\eqref{eq:eta_M}, a sufficient sample size is
\begin{equation}
    M > \frac{2}{\gamma_\varepsilon^2} \log\!\left(\frac{2L_n}{\delta}\right).
    \label{eq:p1_sample_complexity}
\end{equation}

\begin{theorem}[Finite-sample guarantee for P1]\label{thm:P1}
Let $\varepsilon>0$, $\delta\in(0,1)$, and let $\gamma_\varepsilon$ be given by Eq.~\eqref{eq:gamma_epsilon}. If $\gamma_\varepsilon>0$ and
\begin{equation*}
M>\frac{2}{\gamma_\varepsilon^{2}}\log\!\left(\frac{2L_n}{\delta}\right),
\end{equation*}
then the threshold traversal run on the Bell estimators $\hat K_\mu$ returns exactly $\mathcal H_\varepsilon$ with probability at least $1-\delta$.
\end{theorem}

Importantly, no condition is imposed on internal nodes containing no target coefficient. Statistical fluctuations may cause such nodes to be expanded, thereby increasing the running time, without affecting the correctness of the algorithm.

\paragraph*{Top-$t$ identification (P2).}

For P2, the separating value $\theta_t$ in Eq.~\eqref{eq:theta_t} is used only in the analysis and is not an input to the algorithm. Define
\begin{equation}
    \gamma_t
    =
    \min_{\substack{\nu\in\mathcal H_t\\
                     \mu\preceq\nu}}
    \frac{K_\mu-\theta_t}{2^{n-|\mu|}}.
    \label{eq:gamma_t}
\end{equation}
Since every target coefficient satisfies $c_\nu^2\geq w_t>\theta_t$, we have $K_\mu\geq w_t>\theta_t$ for every prefix appearing in Eq.~\eqref{eq:gamma_t}. Hence $\gamma_t>0$ follows automatically from the assumption $w_t>w_{t+1}$.

If
\begin{equation}
    \eta_M(\delta)<\gamma_t,
    \label{eq:p2_condition}
\end{equation}
then Eq.~\eqref{eq:uniform_node_bound} and the definition of $\gamma_t$ imply that every prefix $\mu$ of a target leaf satisfies
\begin{equation}
    |\hat K_\mu-K_\mu|
    \leq 2^{n-|\mu|}\eta_M(\delta)
    < 2^{n-|\mu|}\gamma_t
    \leq K_\mu-\theta_t ,
\end{equation}
so that every prefix containing a target leaf has estimated weight larger than $\theta_t$.

Moreover, the target leaf with squared coefficient $w_t$ is itself included in the minimization in Eq.~\eqref{eq:gamma_t}, and therefore
\begin{equation}
    \gamma_t
    \leq
    w_t-\theta_t
    =
    \frac{w_t-w_{t+1}}{2}.
    \label{eq:gamma_t_leaf_gap}
\end{equation}
Consequently, every non-target leaf $\nu$ satisfies $K_\nu\leq w_{t+1}$, and at a leaf Eq.~\eqref{eq:uniform_node_bound} gives $|\hat K_\nu-K_\nu|\leq\eta_M(\delta)<\gamma_t\leq(w_t-w_{t+1})/2$, so that
\begin{equation}
    \hat K_\nu < w_{t+1} + \frac{w_t-w_{t+1}}{2} = \theta_t.
\end{equation}

As long as at least one target leaf remains unrecovered, one of its prefixes belongs to the current frontier and has estimated weight larger than $\theta_t$. A maximum-weight frontier node therefore cannot be a non-target leaf while an unrecovered target remains. Internal nodes may be selected in any order and may exhibit arbitrary degeneracies; the known identity leaf may also be selected and simply discarded.

Moreover, while an unrecovered target remains, the frontier is nonempty. Since the tree is finite and every node is selected at most once, all target leaves are eventually returned. Consequently, the first $t$ nonidentity leaves returned by the algorithm are exactly $\mathcal H_t$.

Thus P2 succeeds with probability at least $1-\delta$ whenever
\begin{equation}
    M
    >
    \frac{2}{\gamma_t^2}
    \log\!\left(\frac{2L_n}{\delta}\right).
    \label{eq:p2_sample_complexity}
\end{equation}
No uniqueness assumption on intermediate frontier maxima is required.

\begin{theorem}[Finite-sample guarantee for P2]\label{thm:P2}
Let $t\in\mathbb{N}$, $\delta\in(0,1)$, assume $w_t>w_{t+1}$, and let $\gamma_t$ be given by Eq.~\eqref{eq:gamma_t}. If
\begin{equation*}
M>\frac{2}{\gamma_t^{2}}\log\!\left(\frac{2L_n}{\delta}\right),
\end{equation*}
then the first $t$ nonidentity leaves returned by the best-first traversal run on the Bell estimators $\hat K_\mu$ are exactly $\mathcal H_t$, with probability at least $1-\delta$.
\end{theorem}

\paragraph*{Hierarchical versus exhaustive sample requirements.}

Eqs.~\eqref{eq:gamma_epsilon} and
\eqref{eq:gamma_t_leaf_gap} imply
\begin{equation}
    \gamma_\varepsilon
    \leq
    \Delta_\varepsilon,
    \qquad
    \gamma_t
    \leq
    \frac{w_t-w_{t+1}}{2}.
    \label{eq:hierarchical_vs_leaf_margin}
\end{equation}
Thus, at the level of the uniform concentration bounds derived above, the
hierarchical search cannot require fewer Bell samples than exhaustive leaf
processing. Its possible advantage lies instead in reducing the number of
Pauli candidates that must be processed classically. How large the
additional statistical cost is depends on how well the relevant prefix
weights are resolved.

\paragraph*{Pauli-singleton example.}

The distinction between exhaustive leaf estimation and hierarchical search
is already visible for the Pauli-singleton family
\begin{equation}
    \rho_{\mathrm{single}}
    =
    \frac{I^{\otimes n}+X^{\otimes n}}{2^n},
\end{equation}
with $\varepsilon=1/2$. The only nontrivial target is
$X^{\otimes n}$.

For exhaustive processing,
\begin{equation}
    \Delta_\varepsilon
    =
    \frac{1}{4},
\end{equation}
so Eq.~\eqref{eq:exhaustive_p1} requires only
\begin{equation}
    M
    =
    O\!\left(
    n+\log\frac{1}{\delta}
    \right)
\end{equation}
Bell repetitions.

For the hierarchical search, every nonempty target prefix $X^k$ has
$K_{X^k}=1$. At depth one,
\begin{equation}
    \frac{
    K_X-\varepsilon^2
    }{
    2^{n-1}
    }
    =
    \frac{3/4}{2^{n-1}}
    =
    \frac{3}{2^{n+1}}.
\end{equation}
The root has a larger normalized margin, as do the deeper target prefixes,
while $\Delta_\varepsilon=1/4$. Hence, for $n\geq3$,
\begin{equation}
    \gamma_\varepsilon
    =
    \frac{3}{2^{n+1}}.
    \label{eq:singleton_margin}
\end{equation}
The sufficient hierarchical sample bound in
Eq.~\eqref{eq:p1_sample_complexity} therefore scales as
\begin{equation}
    O\!\left(
    4^n
    \left[
    n+\log\frac{1}{\delta}
    \right]
    \right).
\end{equation}

For this family, our sufficient finite-sample guarantee for the hierarchical search is exponentially larger than the corresponding exhaustive guarantee. The singleton is not intrinsically hard for Bell measurements; the cost comes from coarse prefix weights that are small compared with their fluctuations. What the hierarchy gains is classical: the exact traversal evaluates $O(n)$ nodes instead of $4^n$ leaves.

To this end we explore an intermediate approach. We simulate experiments using tolerances for exhaustive search within a hierarchical search. Although this places out of the theoretical guarantees of the method, we conjecture that it lowers the cost of classical post-processing. We examine this conjecture in the next section.

\paragraph*{Structural and computational complexity.}

The statistical guarantees above concern correctness of the returned set.
We now consider the number of tree nodes that must be processed.

For the exact P1 traversal, every expanded internal node satisfies
$K_\mu>\varepsilon^2$. At any fixed depth, the node weights sum to
$N\,\mathrm{Tr}[\aDensityMatrix^2]$, so the number of such nodes is at most
\begin{equation}
    \frac{
    N\,\mathrm{Tr}[\aDensityMatrix^2]
    }{\varepsilon^2}.
\end{equation}
Summing over the $n$ internal levels gives
\begin{equation}
    O\!\left(
    \frac{
    nN\,\mathrm{Tr}[\aDensityMatrix^2]
    }{\varepsilon^2}
    \right)
    \label{eq:p1_structural_bound}
\end{equation}
expanded nodes.

For P2, before the $t$-th target leaf is returned, at least one unrecovered
target lies below a frontier node of weight at least $w_t$. Since the exact
algorithm always selects a maximum-weight node, every internal node expanded
before termination satisfies $K_\mu\geq w_t$. At any fixed depth, at most
\begin{equation}
    \frac{
    N\,\mathrm{Tr}[\aDensityMatrix^2]
    }{w_t}
\end{equation}
nodes can satisfy this condition. Summing over the $n$ internal levels gives
\begin{equation}
    O\!\left(
    \frac{
    nN\,\mathrm{Tr}[\aDensityMatrix^2]
    }{w_t}
    \right)
    \label{eq:p2_structural_bound}
\end{equation}
expanded nodes. Since each expansion generates only four children, the
number of evaluated nodes has the same asymptotic scaling in the exact
traversal.

If the state has only $s$ nonzero nonidentity Pauli coefficients, at most $s+1$ nodes per level have nonzero weight (the ancestors of the $s$ nonzero leaves and of the identity), so the exact traversal expands $O(sn)$ nodes for either problem, irrespective of $N\,\mathrm{Tr}[\aDensityMatrix^2]$. Since $s\log(4^n/s)=s\,(2n\ln 2-\ln s)$, this coincides up to constant factors with the $O(s\log(4^n/s))$ scaling of an adaptive search whenever $s$ is polynomial in $n$; the two differ only when $s$ is exponentially large, in which case neither is polynomial.

The factor $N\,\mathrm{Tr}[\aDensityMatrix^2]$ cannot be avoided for pure states in problem P1. For a pure state, Eq.~\eqref{eq:pauli_norm} gives $\sum_\nu c_\nu^2=2^n$ with $|c_\nu|\leq 1$, so every pure state has at least $2^n$ nonzero Pauli coefficients. Stabilizer states saturate this bound with coefficients of unit magnitude, so for them P1 with any $\varepsilon<1$ necessarily returns $2^n-1$ strings: the exponential cost is intrinsic to the task rather than to the search. For P2 with $t$ polynomial in $n$, Eq.~\eqref{eq:p2_structural_bound} is a worst-case bound, and the number of evaluated nodes can be much smaller when the weight of the $t$ largest coefficients is concentrated on few prefixes. Conversely, purity and statistical conditioning are independent: the Pauli singleton $\rho_{\mathrm{single}}$ has $N\,\mathrm{Tr}[\rho_{\mathrm{single}}^2]=2$, so its exact traversal for $\varepsilon=1/2$ expands at most $8n$ nodes, yet it is statistically hard, as shown above.

These bounds describe the structure of the exact traversal. The finite-sample correctness conditions do not require the sampled traversal
to follow the same sequence of nodes. Statistical fluctuations may cause additional non-target branches to be explored without changing the returned set, so Eqs.~\eqref{eq:p1_structural_bound} and \eqref{eq:p2_structural_bound} should not be interpreted as deterministic
finite-sample running-time bounds.

For the Bell implementation, preprocessing the $M$ measurement records requires $O(Mn)$ operations. Once the run-dependent vector associated with a parent node is available, Eq.~\eqref{eq:child_update} allows all four child vectors to be constructed in $O(M)$ operations. Thus, if $Q$ nodes are evaluated during the actual traversal, the classical post-processing cost is
\begin{equation}
    O(Mn+MQ),
    \label{eq:bell_classical_cost}
\end{equation}
apart from the data structures used to maintain the frontier.

For P1, the order in which nodes above threshold are traversed does not affect either the output or the set of nodes eventually explored. A
depth-first implementation may therefore be used, with no priority queue required. For P2, maintaining the maximum-weight frontier in a priority queue contributes an additional $O(Q\log Q)$ overhead.

The cost in Eq.~\eqref{eq:bell_classical_cost} assumes that the run-dependent vectors needed for future expansions are retained. For P1, a
depth-first implementation need only retain $O(n)$ such vectors at a time, corresponding to $O(Mn)$ storage up to constant factors.

The Bell record itself is collected only once. Since every Bell repetition uses two copies of the state, $M$ repetitions consume $2M$ copies, independently of the number of nodes subsequently evaluated.

For the SWAP implementation, the same fixed-node concentration bound applies, but the measurement circuit depends explicitly on the prefix
$\mu$. For a global guarantee under adaptive node selection, one may conceptually associate an independent $M$-sample record with each possible tree node and apply the union bound before revealing only the records corresponding to queried nodes. Operationally, fresh measurements need only be performed for nodes that are actually queried. If $Q$ nodes are queried, the SWAP implementation therefore requires $O(MQ)$ two-copy experiments rather than a single reusable measurement
record. 

Since $\log L_n=O(n)$, for fixed confidence the hierarchical Bell sample complexity is polynomial whenever the relevant inverse margin,
$\gamma_\varepsilon^{-1}$ for P1 or $\gamma_t^{-1}$ for P2, grows at most polynomially with $n$. A polynomial total running time additionally requires the actual number $Q$ of evaluated nodes to remain polynomial, for example with high probability if a high-probability running-time guarantee is desired. Sample efficiency and computational efficiency are therefore distinct, state-dependent properties of the hierarchical search.





\section{Simulations}\label{sec:simulations}

We evaluate our algorithm on four families of quantum states:
Pauli-singleton states, stabilizer states, transverse-field
Ising ground states, and states with hierarchical Pauli support.
In the figures we report our results for $n=9$ qubits, but for the aggregated comparisons at the end of the section we consider simulations using $n=5$ to $n=10$ qubits.
We compare prefix search with Pauli sampling
\cite{Hinsche2025}, classical shadows \cite{Huang2020} (implemented using both local Pauli and global Clifford measurements), and the adaptive method of Goldar et al.~\cite{goldar2026}.

\subsection{Description of the experiments}

Given an $n$-qubit state $\rho$ and a threshold $\varepsilon$ let 
$$
\mathcal{T}_{\rho}^\varepsilon = \{P \in \{I,X,Y,Z\}^n : \mathrm{Tr}[\rho P]^2 \geq \varepsilon^2 \} \setminus \{I^n\}
$$
denote the set of Pauli matrices whose associated coefficients in $\rho$ have absolute value above $\varepsilon$. Whenever we use one of the algorithms to estimate $\widehat{\mathcal{T}_\rho^\varepsilon}$, we measure the recall of the recovery as
\begin{equation}
    R(\widehat{\mathcal{T}_\rho^\varepsilon},\mathcal{T}_\rho^\varepsilon)
    =
    \frac{|\widehat{\mathcal{T}_\rho^\varepsilon}\cap\mathcal{T}_\rho^\varepsilon|}
         {|\mathcal{T}_\rho^\varepsilon|}.
    \label{eq:simulation_recall}
\end{equation}

For prefix search, we distinguish the top-$t$ variant from the $\varepsilon$ threshold variant. Throughout this section we report the number of state copies consumed, denoted $C$; since each Bell repetition uses two copies, $C$ copies correspond to $M=C/2$ repetitions in the notation of Section~\ref{sec:sample_and_comp}. The parameters and repetitions of each experiment are summarized in
Table~\ref{tab:simulation_parameters}. 

We now briefly describe the other methods implemented.

\begin{table}[t]
    \centering
    \small
    \setlength{\tabcolsep}{3pt}
    \begin{tabular}{lccc}
        \hline
        State
        & $t$
        & $\varepsilon$
        & Reps. (Cliff.) \\
        \hline
        Pauli singleton
        & $1$
        & ---
        & $100$ ($20$) \\
        Stabilizer
        & $2^n-1$
        & $0.5$
        & $100$ ($30$) \\
        Ising
        & $4n+1$
        & $0.6$
        & $100$ ($30$) \\
        Hierarchical
        & $n$
        & $0.9/(2n)$
        & $30$ \\
        \hline
    \end{tabular}
    \caption{
        Parameters used in the simulations. The $t$ and $\varepsilon$ columns give the parameters used for top-$t$ and threshold prefix search, respectively. Parenthesized repetition
        counts correspond to global-Clifford shadows.
    }
    \label{tab:simulation_parameters}
\end{table}

\paragraph{Pauli-sampling.}

Pauli sampling generates candidate Pauli strings sampled according to their Pauli weight by ancestral
sampling from marginals estimated using the same Bell-measurement as prefix search \cite{Hinsche2025}. The number of classical Pauli draws is selected to approximately match
the recovery recall achieved by prefix search. Thus, we are fixing the quality of the recovery of both algorithms, effectively comparing their post-processing performance. We believe this is a reasonable way to compare them considering that they rely on the same set of measurements.

\paragraph{Classical shadows.} We implement the well-known schemes of classical shadows based on Pauli measurements and Clifford measurements~ \cite{Huang2020}. Given a budget of $M$ samples we create a classical shadow of the input state. Later, we query this shadow for every single Pauli matrix, keeping the relevant ones. 

\paragraph{Goldar.} We implement the recent algorithm from \cite{goldar2026} which only works for a certain family of hierarchical states (in which case it recovers them completely). We note that this algorithm cannot be applied to the Pauli-singleton, stabilizer or Ising states. See the hierarchical Pauli support experiment for more details.

\subsection{Pauli-singleton states}

We first consider Pauli-singleton states as
\begin{equation}
    \rho_{\mathrm{single}}
    =
    \frac{I^{\otimes n}+X^{\otimes n}}{2^n}.
    \label{eq:simulation_singleton}
\end{equation}

\begin{figure}[t]
    \centering
    \includegraphics[width=\columnwidth, trim=0 0 0 1.6cm, clip]
        {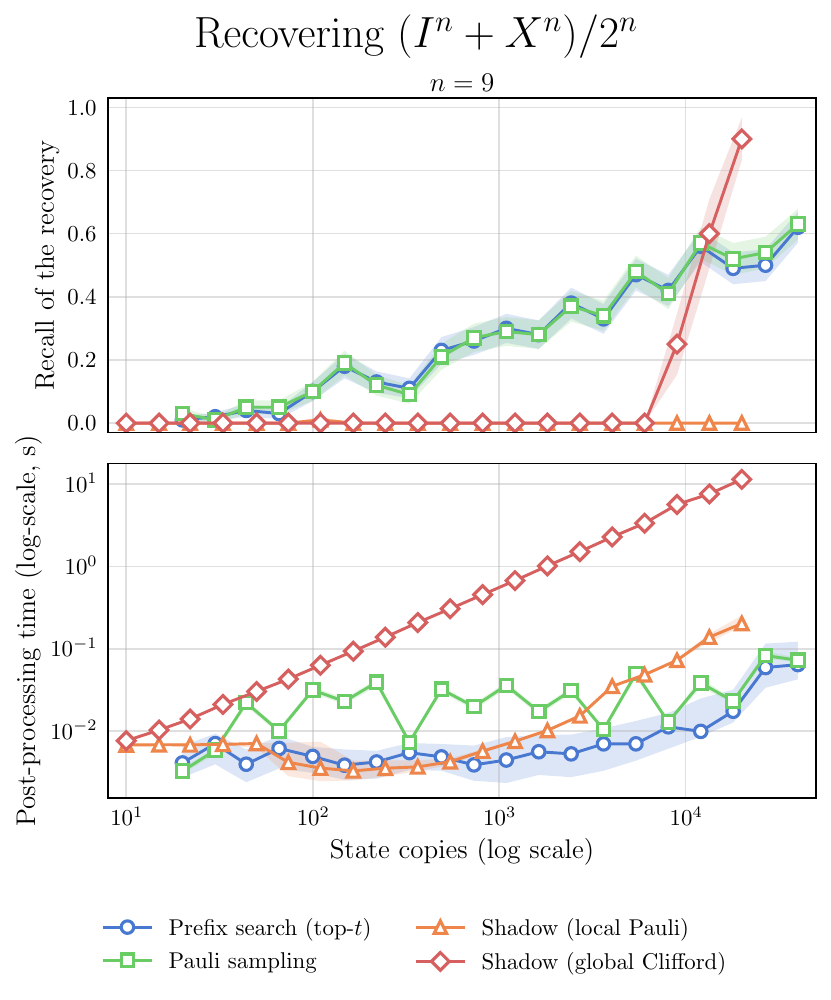}\hfill
    \caption{
        Recall and post-processing time for the Pauli-singleton state
        $(I^{\otimes n}+X^{\otimes n})/2^n$ for $n=9$.
    }
    \label{fig:simulation_singleton}
\end{figure}

Figure~\ref{fig:simulation_singleton} shows that prefix-search success increases with the number of state copies. Despite the sparse support, finite-sample fluctuations in prefix weights remain significant, illustrating that sparsity alone does not ensure reliable discovery with few measurements.

At approximately matched recovery, prefix search generally has a smaller post-processing cost than Pauli sampling. Clifford shadows improve sharply at larger copy budgets and eventually achieve higher success rate, but they require substantially more
post-processing time (and arguably more complex measurements). Pauli shadows have negligible
success over the tested budgets, and since a similar thing happens for the other families of states, we do not include it in the rest of the figures.

\subsection{Stabilizer states}

\begin{figure}[t]
    \centering
    \includegraphics[width=\columnwidth, trim=0 0 0 1.6cm, clip]
        {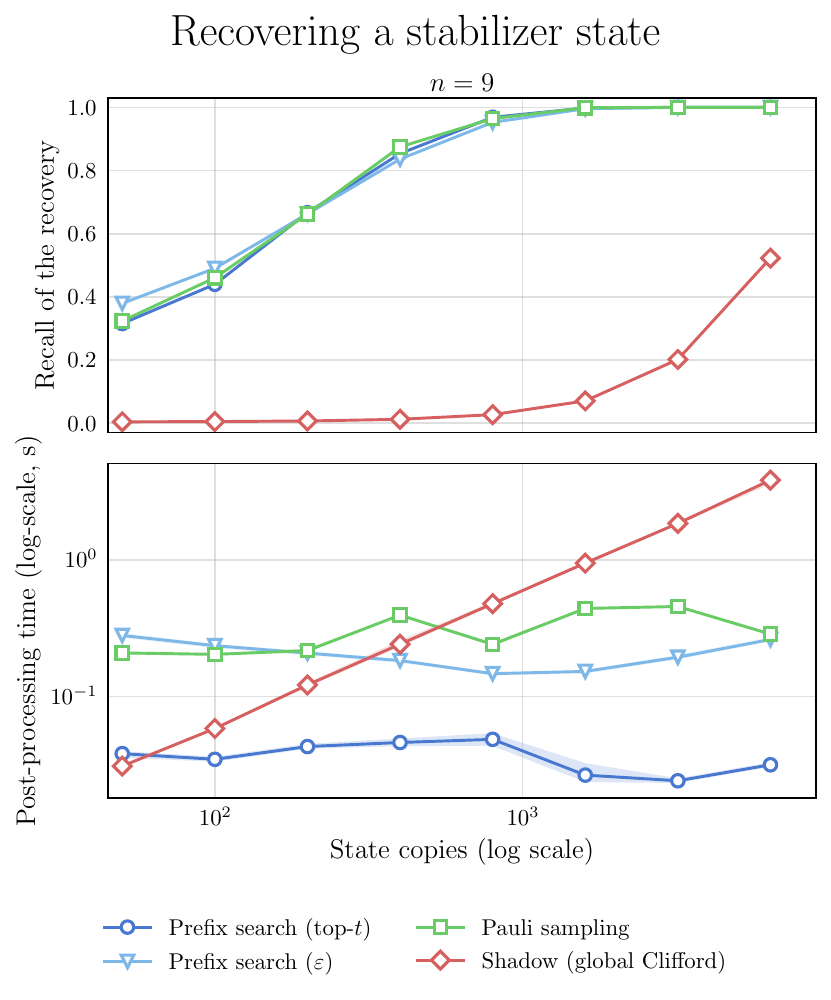}
    \caption{
        Recall and post-processing time for a random stabilizer state of $n$ qubits $n=9$. Upper panels show recovery and
        lower panels show classical post-processing time
        in seconds.
    }
    \label{fig:simulation_stabilizer}
\end{figure}

For each $n$, we generate
\begin{equation}
    |\psi_n\rangle=C_n|0\rangle^{\otimes n},
\end{equation}
where $C_n$ is a random Clifford transformation. One fixed state per $n$ is reused
across all measurement budgets and repetitions. The target set consists of all $2^n-1$ non-identity Pauli
operators with nonzero coefficients, each having
$|c_P|=1$. We use $t=2^n-1$ for top-$t$ search and
$\varepsilon=1/2$ for threshold search.

Figure~\ref{fig:simulation_stabilizer}
shows that both prefix-search variants approach complete
recovery as the copy budget increases. Pauli sampling
achieves similar recall under the calibration procedure,
while global Clifford shadows recover a smaller fraction
of the target support at the same displayed copy budgets.

Top-$t$ prefix search has the smallest post-processing time
among the displayed methods in this benchmark.
Threshold search and Pauli sampling have comparable costs,
whereas the processing time of global Clifford shadows
grows markedly with the number of copies.

\subsection{Transverse-field Ising ground states}

We consider the periodic transverse-field Ising Hamiltonian
\begin{equation}
    H =
    -J\sum_{i=1}^{n}X_iX_{i+1}
    -h\sum_{i=1}^{n}Z_i,
    \qquad
    X_{n+1}=X_1,
    \label{eq:simulation_ising}
\end{equation}
with $J=h=1$. The ground state is obtained by numerical
diagonalization.

We consider the target set
\begin{equation}
    \mathcal{T}_{\mathrm{Ising}}
    =
    \left\{
        P\neq I^{\otimes n}:
        |\langle\psi_0|P|\psi_0\rangle|>0.6
    \right\}.
\end{equation}
Direct enumeration gives $4n+1$ target operators for the
tested sizes $n=5,\ldots,10$. We therefore set $t=4n+1$ for top-$t$ search and use
$\varepsilon=0.6$ for the threshold variant.

\begin{figure}[t]
    \centering
    \includegraphics[width=\columnwidth, trim=0 0 0 1.6cm, clip]
        {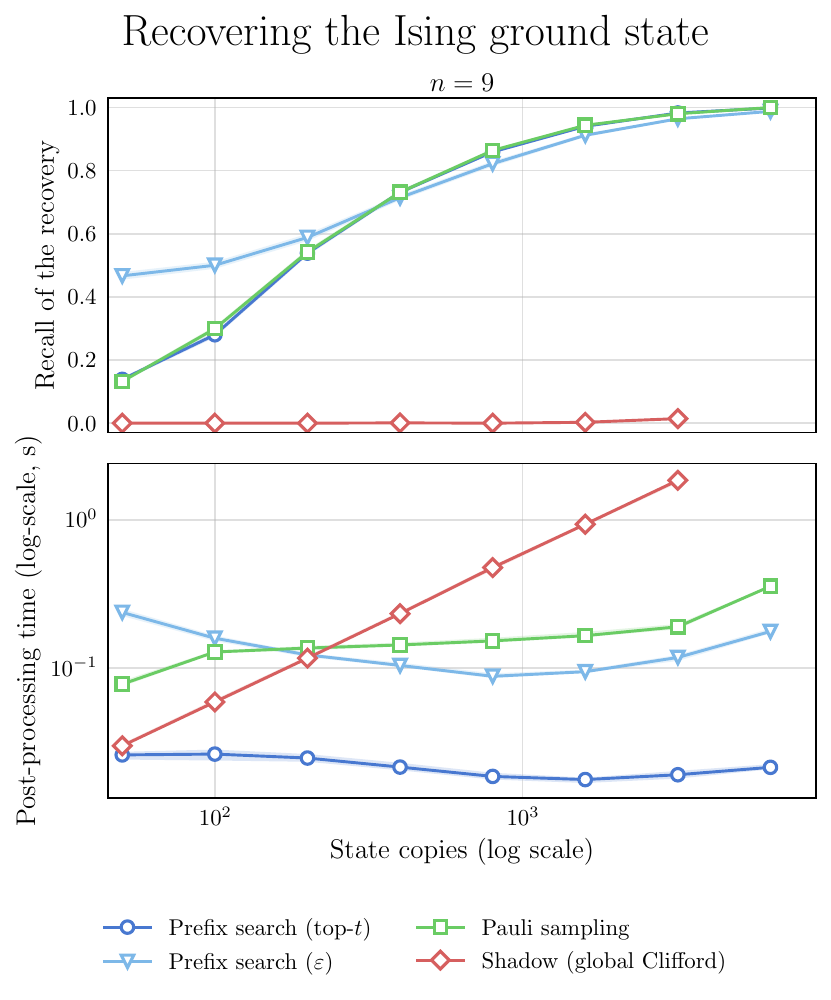}
    \caption{
        Results regarding the recovery of the ground state
        of the periodic transverse-field Ising Hamiltonian
        in Eq.~\eqref{eq:simulation_ising}, with $J=h=1$,
        for $n=9$.
        Top-$t$ prefix search uses $t=4n+1$, while threshold
        search uses $0.6$ as threshold.
    }
    \label{fig:ising}
\end{figure}

Figure~\ref{fig:ising} shows results similar to those of the stabilizer case: the top-$t$ prefix search algorithm is the most efficient in terms of post-processing time, being up to 10 times faster than the other variant of prefix search and Pauli sampling.

\subsection{States with hierarchical Pauli support}

Finally, we consider
\begin{equation}
    \rho_\beta
    =
    \frac{1}{2^n}
    \left(
        I^{\otimes n}
        +
        \beta\sum_{j=1}^{n}
        X^{\otimes j}\otimes I^{\otimes(n-j)}
    \right),
    \label{eq:simulation_hierarchical}
\end{equation}
with $\beta=0.9/n$. For such a family of states the adaptive algorithm from \cite{goldar2026} allows for efficient (i.e. polynomial time) and complete recovery. The targets are the $n$ non-identity prefix operators
$X^{\otimes j}\otimes I^{\otimes(n-j)}$.
We use $t=n$ for top-$t$ search and
$\varepsilon=\beta/2$ for threshold search.

\begin{figure}[t]
    \centering
    \includegraphics[width=\columnwidth, trim=0 0 0 1.5cm, clip]
        {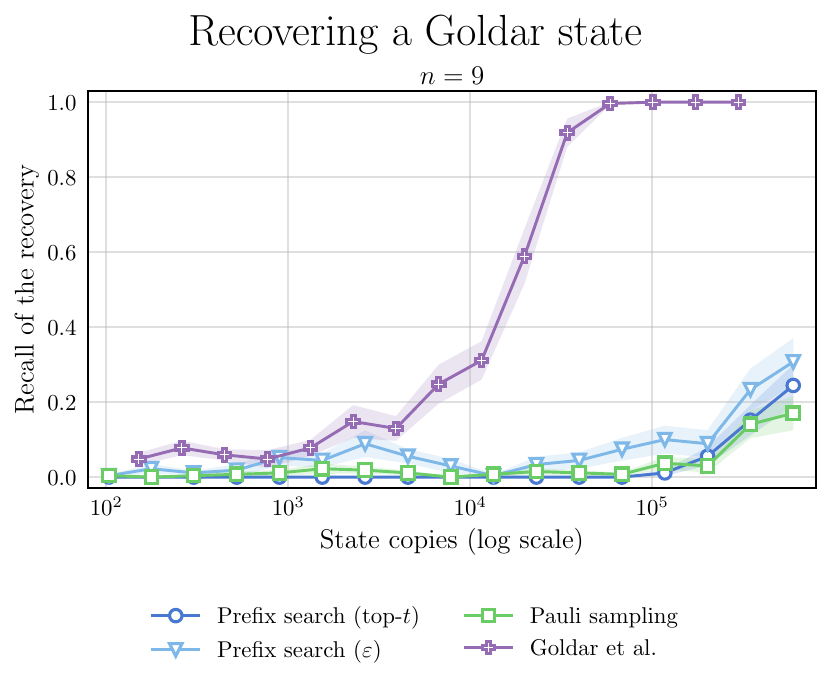}
    \caption{
        Recall for the hierarchical state
        $\rho_\beta$ in
        Eq.~\eqref{eq:simulation_hierarchical},
        with $\beta=0.9/n$.
        The targets are the $n$ non-identity prefix
        operators. Top-$t$ prefix search uses $t=n$,
        and threshold search uses
        $\theta=(\beta/2)^2$.
    }
    \label{fig:goldar}
\end{figure}

Figure~\ref{fig:goldar} shows that the adaptive method reaches
nearly complete recovery at substantially smaller copy
budgets than prefix search and fixed-budget Pauli sampling.
The threshold version of prefix search generally has higher
recall than the top-$t$ version in this experiment, although
both remain below the adaptive benchmark over the displayed
range.

\subsection{Discussion}

We now summarize the main observations from the numerical experiments. We first compare the methods in terms of sample efficiency and classical post-processing cost, and then discuss the role of additional structural information and the dependence of processing time on the number of copies.

\paragraph{Sample complexity.}

For stabilizer and Ising states our algorithm outperformed the Clifford Shadow approach in terms of sample complexity, while in the case of Pauli-singleton states the latter algorithm seems to be better. In Table~\ref{tab:sample_complexity_comparison} reports the number of samples required by each algorithm to achieve a recovery of 70\% for some families of states. Note that top-$t$ prefix search reaches this target with
fewer state copies than the global-Clifford classical-shadow for the more realistic states, and that the gap seems to increase as $n$ grows. In particular, for Ising states the classical shadows algorithm doesn't even achieve that level of recovery with the tested budget. 

Also, observe that for both stabilizer and Ising states the top-$t$ search achieves a good recall with very few samples. This is surprising considering that the sample complexity bound obtained in Section~\ref{sec:sample_and_comp} is exponential in general.

\begin{table}[tb]
    \centering
    \setlength{\tabcolsep}{4pt}
    \renewcommand{\arraystretch}{1.15}
    \begin{tabular}{@{}l@{\hspace{10pt}}crrc@{}}
        \cline{2-5}
        & $n$ & Top-$t$ & Shadows & Ratio \\
        \cline{2-5}
        \multirow{6}{*}{\shortstack[l]{Pauli-\\singleton}}
        & $4$ & $95$       & $212$     & $2.2\times$ \\
        & $5$ & $271$      & $612$     & $2.3\times$ \\
        & $6$ & $696$      & $1064$    & $1.5\times$ \\
        & $7$ & $2294$     & $3533$    & $1.5\times$ \\
        & $8$ & $12\,046$  & $7195$    & $0.6\times$ \\
        & $9$ & $>40\,000$ & $15\,318$ & --- \\
        \cline{2-5}
        \multirow{5}{*}{Stabilizer}
        & $5$ & $59$  & $341$   & $5.8\times$ \\
        & $6$ & $112$ & $774$   & $6.9\times$ \\
        & $7$ & $78$  & $1798$  & $22.9\times$ \\
        & $8$ & $380$ & $3980$  & $10.5\times$ \\
        & $9$ & $235$ & $>6400$ & --- \\
        \cline{2-5}
    \end{tabular}
    \caption{
        Number of copies required (on average) to achieve $70\%$ recovery using top-$t$ prefix search versus global Clifford classical shadows. The ratio column indicates the improvement of top-$t$ over classical shadows (a value above 1 indicates that top-$t$ outperforms the latter). We do not show the entry for $n=10$ since, as it happens with $n=9$, one of the two algorithms never reaches 70\% recall.
    }
    \label{tab:sample_complexity_comparison}
\end{table}
\paragraph{Classical post-processing performance} Across all four state families, top-$t$ prefix search achieves a
notable post-processing speedup over Pauli sampling, the
next-best competing method in our post-processing benchmarks.
Table~\ref{tab:postprocessing_speedups} summarizes the median
speedup factors, with values greater than one indicating
faster processing by prefix search. This speed-up is reasonable, considering in particular that both algorithms exploit the tree-structure of the Pauli prefixes, but ours is specifically designed to identify the largest coefficients in the Pauli basis.

\begin{table}[tb]
    \centering
    \begin{tabular}{lc}
        \hline
        State family & Median speedup \\
        \hline
        Pauli singleton    & $1.68\times$ \\
        Stabilizer         & $5.17\times$ \\
        Ising ground state & $7.07\times$ \\
        \hline
    \end{tabular}
    \caption{
        Median classical post-processing speedup of prefix
        search relative to Pauli sampling across the
        evaluated settings, considering all values of $n$ from $5$ to 10. 
    }
    \label{tab:postprocessing_speedups}
\end{table}

\paragraph{Hierarchical Pauli support.}

On the hierarchical family in Eq.~\eqref{eq:simulation_hierarchical}, the adaptive method of Goldar et al.~\cite{goldar2026} achieves better sample performance than all other methods tested. This is reasonable considering that this algorithm takes advantage of prior knowledge of the state family. In particular, this strategy cannot be applied to other families of states, while prefix search, in contrast, does not rely on any hypothesis (although its performance, as we can see, varies from family to family).

\paragraph{Post-processing time does not seem to scale with the number of copies}

The figures show that for both prefix search and Pauli sampling post-processing time varies only weakly with the number of copies. Although the cost of evaluating an individual node of the tree increases with the
number of copies, each node weight is computed only once for a given set of measurement results because we cache these intermediate values. Thus, subsequent visits to the same node reuse its stored value. This is particularly relevant for Pauli sampling, since it traverses the tree as many times as the number of classical samples requested. At the same time, a larger number of copies reduces statistical fluctuations, making the underlying Pauli structure more accurately reflected in the estimated node weights, therefore reducing the amount of overall nodes traversed by prefix search. Meanwhile, classical shadow post-processing time always scales with the number of copies because the size of the classical shadow is proportional to it.

\section{Conclusion}\label{sec:conclu}

In this work we developed a hierarchical algorithm for identifying the largest Pauli coefficients of a quantum state. The novelty of the proposed algorithm lies in the fact that it identifies the largest coefficients first without the need to resort to full tomography. Moreover, the algorithm is flexible and can be adapted to other types of partitions for node expansion beyond simply appending a character to a prefix.



Its cost is governed by two independent factors. The first is statistical: the Bell estimator of a prefix weight has a fluctuation scale $2^{n-|\mu|}/\sqrt{M}$, so the traversal returns the correct set only once the number of samples resolves the state-dependent margins $\gamma_\varepsilon$ or $\gamma_t$ of Theorems~\ref{thm:P1} and \ref{thm:P2}. These margins are never larger than the corresponding leaf separations, so the uniform sufficient sample bounds derived here are never smaller than those for exhaustive processing of the same Bell record; the potential advantage of the hierarchical strategy is instead classical, in the number of Pauli candidates that must be examined. The second factor is structural: the number of nodes explored in the exact traversal, bounded by Eqs.~\eqref{eq:p1_structural_bound} and \eqref{eq:p2_structural_bound}, is small only when the relevant weight is concentrated on a few prefixes.

The numerical experiments illustrate both regimes. Pauli-singleton states have a tiny relevant subtree but exponentially diluted prefix weights, and on them the method is outperformed by Clifford shadows at large $n$. For the stabilizer and transverse-field Ising instances tested, top-$t$ prefix search reaches a given recall with fewer copies than global-Clifford classical shadows and with a post-processing time several times smaller than that of Pauli sampling on the same data. This favorable behavior is consistent with the relevant prefix weights being sufficiently well separated to be resolved at the explored system sizes. On states with hierarchical Pauli support, the adaptive method of Ref.~\cite{goldar2026}, which exploits a promise on the family, needs far fewer copies, whereas prefix search applies without assumptions.

Finally, it is worth noting that certain symmetries of the states can help identify coefficients. For example, if the outcome for $X$ on the first component is always positive, the only Bell-basis outcomes compatible with this imply that $Z$ and $Y$ have opposite signs. This leads to the estimators of the corresponding nodes summing to zero. Since the corresponding node-weight estimators sum to zero for every measurement record, their expectations must also sum to zero. As the exact node weights are nonnegative, each of these weights must vanish. Consequently, all coefficients belonging to the corresponding branches vanish.

\paragraph*{Acknowledgements}

This work was possible despite the lack of support from the scientific funding agencies of Argentina.

\appendix

\section{Direct derivation of the Bell node estimator}
\label{app:bell_estimator}
 
In the main text, Eq.~\eqref{eq:bell_K_estimator} was obtained from the
operator representation of the node weights. Here we give an alternative
derivation directly from the sign patterns associated with Bell-measurement
outcomes.
 
For a Pauli string $\nu$, a Bell-measurement run $j$ determines a sign
$s_j(P_\nu\otimes P_\nu)\in\{-1,+1\}$ satisfying
\begin{equation}
    \mathbb E\!\left[
    s_j(P_\nu\otimes P_\nu)
    \right]
    =
    c_\nu^2 .
\end{equation}
Hence, for a prefix $\mu$,
\begin{equation}
    \hat K_\mu
    =
    \sum_{\nu\in\Gamma_\mu}
    \frac{1}{M}
    \sum_{j=1}^{M}
    s_j(P_\nu\otimes P_\nu).
    \label{eq:app_node_sum}
\end{equation}
 
Let $k=|\mu|$. Writing $\nu=\mu\beta$, where $\beta$ is a suffix of
length $n-k$, the sign factorizes as
\begin{equation}
    s_j(P_\nu\otimes P_\nu)
    =
    s_j(\mu)\,s_j(\beta),
\end{equation}
where $s_j(\mu)$ is the product of the signs associated with the first
$k$ qubit pairs. Interchanging the order of summation in
Eq.~\eqref{eq:app_node_sum}, we obtain
\begin{equation}
    \hat K_\mu
    =
    \frac{1}{M}
    \sum_{j=1}^{M}
    s_j(\mu)\,
    \underbrace{
    \left(
    \sum_{\beta}s_j(\beta)
    \right)
    }_{S_{\mu,j}} .
    \label{eq:app_node_factorized}
\end{equation}
 
To evaluate $S_{\mu,j}$, we count the suffix paths that produce positive
and negative total signs. For a fixed run $j$, let
$\mathrm{paths}^+(r,j)$ and $\mathrm{paths}^-(r,j)$ denote the number of
Pauli suffixes beginning at position $r$ whose accumulated sign is,
respectively, positive and negative. We define
\begin{equation}
    \Delta\mathrm{paths}(r,j)
    =
    \mathrm{paths}^+(r,j)
    -
    \mathrm{paths}^-(r,j).
    \label{eq:app_delta_paths}
\end{equation}
Then
\begin{equation}
    S_{\mu,j}
    =
    \Delta\mathrm{paths}(k,j).
\end{equation}
 
For each Bell outcome, the signs of
$(X,Y,Z,I)$ are one of
\begin{equation}
    {-}+++,
    \qquad
    +{-}++,
    \qquad
    ++{-}+,
    \qquad
    {-}{-}{-}+,
    \label{eq:app_bell_patterns}
\end{equation}
corresponding to $|\Phi^-\rangle$, $|\Phi^+\rangle$, $|\Psi^+\rangle$ and the
singlet $|\Psi^-\rangle$, respectively.
For the first three patterns, three of the four possible Pauli choices
preserve the sign of the remaining path and one reverses it. Therefore
\begin{align}
    \mathrm{paths}^+(r,j)
    &=
    3\,\mathrm{paths}^+(r+1,j)
    +
    \mathrm{paths}^-(r+1,j),
    \\
    \mathrm{paths}^-(r,j)
    &=
    3\,\mathrm{paths}^-(r+1,j)
    +
    \mathrm{paths}^+(r+1,j),
\end{align}
and hence
\begin{equation}
    \Delta\mathrm{paths}(r,j)
    =
    2\,\Delta\mathrm{paths}(r+1,j).
\end{equation}
 
For the singlet pattern ${-}{-}{-}+$, three choices reverse the sign and
only one preserves it, giving
\begin{equation}
    \Delta\mathrm{paths}(r,j)
    =
    -2\,\Delta\mathrm{paths}(r+1,j).
\end{equation}
 
Introducing
\begin{equation}
    a_{r,j}
    =
    \begin{cases}
        1, & \text{if pair $r$ is the singlet},\\
        0, & \text{otherwise},
    \end{cases}
\end{equation}
both cases can be written as
\begin{equation}
    \Delta\mathrm{paths}(r,j)
    =
    (-1)^{a_{r,j}}
    2\,
    \Delta\mathrm{paths}(r+1,j).
    \label{eq:app_delta_recursion}
\end{equation}
 
The recursion is initialized beyond the last qubit, where there is a single
empty suffix with positive sign:
\begin{equation}
    \Delta\mathrm{paths}(n,j)=1.
\end{equation}
Iterating Eq.~\eqref{eq:app_delta_recursion} backwards gives
\begin{equation}
    \Delta\mathrm{paths}(k,j)
    =
    2^{n-k}
    (-1)^{m_{k,j}},
    \qquad
    m_{k,j}
    =
    \sum_{r=k}^{n-1}a_{r,j}.
    \label{eq:app_delta_closed}
\end{equation}
Only the parity of $m_{k,j}$ enters the expression.
 
Substituting Eq.~\eqref{eq:app_delta_closed} into
Eq.~\eqref{eq:app_node_factorized}, we recover
\begin{equation}
    \hat K_\mu
    =
    \frac{2^{n-k}}{M}
    \sum_{j=1}^{M}
    s_j(\mu)(-1)^{m_{k,j}},
\end{equation}
which is Eq.~\eqref{eq:bell_K_estimator} of the main text.

\bibliographystyle{apsrev4-2}
\bibliography{bibliography}

\end{document}